\documentclass[fleqn,usenatbib]{mnras}

\usepackage{newtxtext,newtxmath}

\usepackage[T1]{fontenc}
\usepackage{ae,aecompl}

\usepackage{graphicx}	
\usepackage{amsmath}	
\usepackage{amssymb}	

\title[A general-purpose gravitational timestep]{A general-purpose timestep criterion for simulations with gravity}

\author[Grudi\'{c} \& Hopkins]{
Michael Y. Grudi\'{c}$^{1,2}$\thanks{E-mail: mike.grudic@northwestern.edu} and Philip F. Hopkins$^{2}$
\\
$^{1}$Department of Physics and Astronomy and CIERA, Northwestern University, 2145 Sheridan Road, Evanston, IL 60208, USA\\
$^{2}$TAPIR, Mailcode 350-17, California Institute of Technology, Pasadena, CA 91125, USA\\
}

\date{Accepted XXX. Received YYY; in original form ZZZ}

\pubyear{2018}

\begin{document}
\label{firstpage}
\pagerange{\pageref{firstpage}--\pageref{lastpage}}
\maketitle

\begin{abstract}
We describe a new adaptive timestep criterion for integrating gravitational motion, which uses the tidal tensor to estimate the local dynamical timescale and scales the timestep proportionally. This provides a better candidate for a truly general-purpose gravitational timestep criterion than the usual prescription derived from the gravitational acceleration, which does not respect the equivalence principle, breaks down when $\mathbf{a}=0$, and does not obey the same dimensional scaling as the true timescale of orbital motion. We implement the tidal timestep criterion in the simulation code {\small GIZMO}, and examine controlled tests of collisionless galaxy and star cluster models, as well as galaxy merger simulations. The tidal criterion estimates the dynamical time faithfully, and generally provides a more efficient timestepping scheme compared to an acceleration criterion. Specifically, the tidal criterion achieves order-of-magnitude smaller energy errors for the same number of force evaluations in potentials with inner profiles shallower than $\rho \propto r^{-1}$ (ie. where $\mathbf{a}\rightarrow 0$), such as star clusters and cored galaxies. For a given problem these advantages must be weighed against the additional overhead of computing the tidal tensor on-the-fly, but in many cases this overhead is small.
\end{abstract}



\section{Introduction}
Numerical astrophysical simulations nearly always involve (1) gravity, and (2) a wide dynamic range of spatial and temporal scales. When gravity is important, simulations must evaluate the gravitational field at every point in the simulation domain, typically an expensive operation, but in many cases this only needs to be done frequently for a small subset of the simulation that is evolving on short timescales. The use of an {\it adaptive} timestep criterion can permit calculations that are otherwise computationally prohibitive, if it can be properly scaled to the physical timescale of the system. Even if the simulation advances on a ``uniform'' timestep $\Delta t$ (all elements advancing simultaneously), this timestep is almost always adaptive {\em in time}, usually chosen to be the largest possible timestep which satisfies some stability or accuracy criterion for all resolution elements at a given instant.

Achieving both numerical stability and accuracy requires, in general, that $\Delta t$ is smaller than the timescales over which the system is dynamically changing. For e.g.\ a test particle in a circular orbit (velocity $V_{c}$) at a radius $r$ around a point mass of mass $M_{\rm enc}$, this is just the usual dynamical time 
\begin{equation}
    t_{\rm dyn} = \Omega^{-1} = \frac{r}{V_{c}} = \sqrt{\frac{r^3}{G M_{\rm enc}}},
    \label{eq:tdyn}
\end{equation}
which would provide a natural candidate for the timestep 
\begin{equation}
    \Delta t \leq \sqrt{\eta}\,t_{\rm dyn},
\end{equation}
(where $\eta$ is a dimensionless tolerance parameter that sets the overall integration accuracy, dependent on the actual integration method). Of course, in this case, the relevant timescale is obvious precisely because the simplicity of the problem means analytic solutions are known, which would allow us to use arbitrarily large timesteps. The challenge in e.g.\ N-body simulations is deriving a truly general criterion when the solutions are not known (hence the simulation).


So-called ``collisional'' or ``direct'' N-body codes \citep[e.g.][]{nbody6} designed to integrate the trajectories of a (relatively small) number of point masses can resolve this by using high-order integration methods for which accurate higher-order derivatives of the particle motion are available. Different ratios of these derivatives can then be combined into satisfactory timestep criteria \citep{aarseth:2003}. 
However, many gravity codes designed for cosmology, galactic dynamics, star formation or star cluster dynamics, or small-body (e.g.\ asteroid, dust) planetary dynamics, as well as multi-physics applications such as fluid dynamics or stellar structure, are necessarily ``approximate.'' These may feature ``N'' bodies much less than the true number of ``particles'' or ``point masses,'' self-gravity of continuous fields like fluids, softened gravity, approximate (but efficient) representations of long-range forces (e.g.\ tree or multipole or particle-mesh methods), or a wide variety of {\em other} timestep constraints (e.g.\ the Courant condition) which must also be considered. This generally bars the use of such high-order integration methods. In these applications, the most widely-adopted solution is to construct a timescale from the acceleration $\mathbf{a}$ and the ``size'' $\epsilon$ of the resolution element:
\begin{equation}
    \Delta t \leq \sqrt{\frac{\eta \epsilon}{ |\mathbf{a}|}},
    \label{eq:power}
\end{equation}
where $\epsilon$ has dimensions of length, and could correspond to the actual hydrodynamic cell size for an Eulerian grid code \citep[e.g.][]{art, orion, athena, flash, ramses, enzo}, moving mesh code \citep{springel:arepo, tess, disco}, or meshless Arbitrary Lagrangian Eulerian (ALE) code \citep{hopkins:gizmo, gandalf}, the kernel size for a smoothed particle hydrodynamics code \citep{rasio:1991.sph, monaghan:1992.sph, bate:1995, springel:gadget,gasoline2, phantom}, or the simply the gravitational softening length for N-body simulations. We will refer to Eq.~\ref{eq:power} as the ``acceleration criterion". \citet{power:2003} showed that this timestep criterion gives satisfactory results in dark matter-only cosmological simulations, adequately controlling artificial heating due to integration errors and thus ensuring that the accuracy of the simulation was limited by other factors such as resolution and force accuracy. It is the de-facto standard timestep criterion for galaxy simulations in particular, but has also found use in many other types of simulations, and is implemented in many of the codes cited above.

Although it does perform well in certain applications, Eq.~\ref{eq:power} is not wholly satisfactory as a truly {\it general-purpose} gravitational timestep criterion. One issue is that it does not respect an important physical symmetry. Consider a system embedded in an externally-imposed uniform gravitational field $\mathbf{g}_{\rm ext}$. $\mathbf{g}_{\rm ext}$ can be arbitrary large, to the point of dominating $\mathbf{a}$, but this does not alter the equations of motion for the system's internal dynamics, once transformed to the uniformly-accelerated frame. Moreover, any integration scheme of second order or higher (i.e.\ any scheme used for serious numerical calculations) integrates the uniformly-accelerated component of the motion {\it exactly}, so it should not enter the timestep criterion. However Eq.~\ref{eq:power} {\it would} be altered, and the numerical solution in turn. 

Eq.~\ref{eq:power} also fails to accurately capture the timescale of motion in arbitrary systems. For instance, if integrating motion of the Earth-Moon system, clearly $\Delta t$ ought to be some fraction of a month. However, once one considers the Sun, the $\mathbf{a}$ experienced by the Earth and Moon is dominated by the contribution of the Sun's gravitational field, which is two orders of magnitude greater than the fields exerted by the Earth upon the Moon, so the acceleration criterion would ignore the details of the Earth-Moon system. Moreover in an $N$-body system with finite $\epsilon$ introduced around e.g.\ some ``stars'' or ``planetary bodies'' (to represent either their true physical extent or to simply prevent divergent accelerations in close passages), clearly $\epsilon$ has no physical significance in the gravitational orbits when it is much smaller than the separation between those bodies. Eq.~\ref{eq:power} can also fail catastrophically in certain instances. Consider a spherically-symmetric potential with finite central density $\rho_0$: test-particle orbits about the center can be approximated to leading order as simple harmonic motion with period 
\begin{equation}
T_0 =\sqrt{\frac{3\pi }{G \rho_0}},
\end{equation}
but Eq.~\ref{eq:power} gives a timestep 
\begin{equation}
    \Delta t \leq \sqrt{\frac{\eta \epsilon}{r}}\, T_0,
\end{equation}
which diverges as $r\rightarrow 0$ (some other timestep limit will explicitly or implicitly be needed), and so particles can take artificially large ``jumps'' if their timestep is evaluated when $r\rightarrow 0$. Even in the opposite limit (``large'' $r\gg \epsilon$), this is inefficient, demanding a number of timesteps per orbit $\sim (r/\epsilon)^{1/2}$ despite $\epsilon$ playing no role in the dynamics here.\footnote{Of course many other timestep criteria have also been proposed, which have similar issues to Eq.~\ref{eq:power} and are less widely-used, including Courant-like criteria $\Delta t \propto \epsilon/| \delta {\bf v}|$ (where $\delta{\bf v}$ is a local relative/signal velocity) or $\Delta t \propto |\delta {\bf v}|/|{\bf a}|$, or local self-gravity criteria with $\Delta t \propto (G\rho)^{-1/2}$ \citep[see discussion in][]{power:2003,fire2}. The former break down like Eq.~\ref{eq:power} at orbital apocenters, Lagrange points, and/or potential extrema, and also do not respect the equivalence principle; the latter cannot account for any external or collective forces.}


\citet{zemp:2007} noted this and proposed a solution wherein $\Delta t$ is scaled to an estimate of $t_{\rm dyn}$ in Eq.~\ref{eq:tdyn} obtained by determining some ``effective enclosed mass and radius'' about which each particle is orbiting (calculating an ``effective density $M_{\rm enc}/R^{3}$ between each pair of particles and total mass enclosed in various directional ``cones''). While is more general than Eq.~\ref{eq:power}, it still remains closely tied to the idea of quasi-circular orbits in spherically-symmetric potentials and. 

But there is a conceptually simple and well-motivated local estimator of $t_{\rm dyn}$ which does not require arbitrary parameters, specific assumptions about the geometry or matter distribution or types of orbits, or added computational complexity (processing of ``interaction lists'' or many passes through the gravity tree). Consider the tidal tensor $\mathbf{T}$, the gradient of the gravitational field:
\begin{equation}
    \mathbf{T} \equiv \mathbf{\nabla\,g}.
\end{equation}
$\mathbf{T}$ has dimensions of $[\mathrm{time}]^{-2}$, and can directly be converted to a timescale, is independent of uniform external accelerations, is well-defined for any orbit structure or matter distribution, is computable for any method with a well-defined ${\bf g}$, and scales in a manner similar to Eq.~\ref{eq:tdyn} for spherically-symmetric systems (as desired). In fact \citet{dehnen:2011.review} did consider a timestep criterion derived from $\mathbf{T}$, noting these advantages as well as the fact that it captured the desired behaviors of higher-order estimators described above, but dismissed the computation of $\mathbf{T}$ as computationally ``extravagant.'' While it may be true that for some specialized N-body applications the additional overhead is significant, in many modern memory-intensive multi-physics simulations the overhead of computing $\mathbf{T}$ is negligible \citep[and in many simulations $\mathbf{T}$ is already computed for other purposes;][]{vogelsberger:2008.gde, renaud:2017.cluster.zooms, pfeffer:2018.emosaics, li.gnedin:2019.cluster.zooms}. In this paper, we will show that $\mathbf{T}$ can provide an excellent estimator of the local dynamical time and a simple, flexible adaptive timestep criterion with a number of advantages (at least in some problems) compared to previous prescriptions.

\section{A Tidal Timestep Criterion}
\subsection{Definition}
To convert $\mathbf{T}$ to a timestep, we consider the tidal tensor in a Keplerian potential:
\begin{equation}
    T_{ij}^{K} = \frac{GM}{r^3}\left(-\delta_{ij} + \frac{3 x_i x_j}{r^2}\right),
    \label{eq:tij_kep}
\end{equation}
where $x_i$ are the Cartesian coordinates and $r^2=\sum x_{i}^{2}$ is the distance from the central mass $M$ (defined as the origin). Obviously $\mathbf{T}$ encodes the dynamical time $t_{\rm dyn}=\sqrt{r^3/GM}$, but we desire a positive-definite, coordinate-invariant scalar to determine the timestep, which can be constructed from its eigenvalues\footnote{Since $\mathbf{T}$ is a real symmetric tensor, it has 3 real $\lambda_i$.} $\lambda_i$. A natural candidate is the invariant Frobenius norm:\footnote{The invariant ${\rm Trace}(\mathbf{T}) = \nabla \cdot {\bf g} = -\nabla^{2}\phi = -4\pi\,G\,\rho$ is not useful for timesteps as it contains only the local density information (no long range forces) and vanishes in vacuum. The determinant of $\mathbf{T}$ can be used to define a timestep $\Delta t = (|{\rm det}(\mathbf{T})|/2)^{-1/6}$ which behaves similarly to our Frobenius norm prescription for spherical potentials, but $\mathbf{T}$ becomes ill-conditioned (and the timestep vanishes) if the matter distribution is highly anisotropic (in e.g.\ sheets or filaments).}
\begin{equation}
    ||\mathbf{T}||^2 = \sum_i^3 \lambda_i^2 = \sum_i^3 \sum_j^3 T_{ij}^2,
\end{equation}
where the latter expression holds in any Cartesian coordinate system, and is convenient because it does not require the values of $\lambda_i$ to be computed explicitly. Computing this for the Keplerian tidal tensor (Eq.~\ref{eq:tij_kep}), we see that
\begin{equation}
    ||\mathbf{T}^K||^2 = \frac{6 G^2 M^2}{r^6}.
\end{equation}
Thus, we define the invariant {\it tidal timescale} $t_{\rm tidal}$ such that the dynamical time is recovered in the Keplerian limit:
\begin{equation}
    t_{\rm tidal} =\left(\frac{||\mathbf{T}||^2}{6}\right)^{-1/4}.
\end{equation}
Then, the tidal timestep criterion is that the timestep be less than some fraction of $t_{\rm tidal}$:
\begin{equation}
    \Delta t \leq \sqrt{\eta}\,t_{\rm tidal},
    \label{eq:dt_tidal}
\end{equation}
keeping the convention where $\eta$ is proportional to the integration error for a second-order integrator. This defines the tidal timestep criterion.

\subsection{General behaviour and comparison with acceleration criterion}
\begin{figure*}
    \centering
    \includegraphics[width=\textwidth]{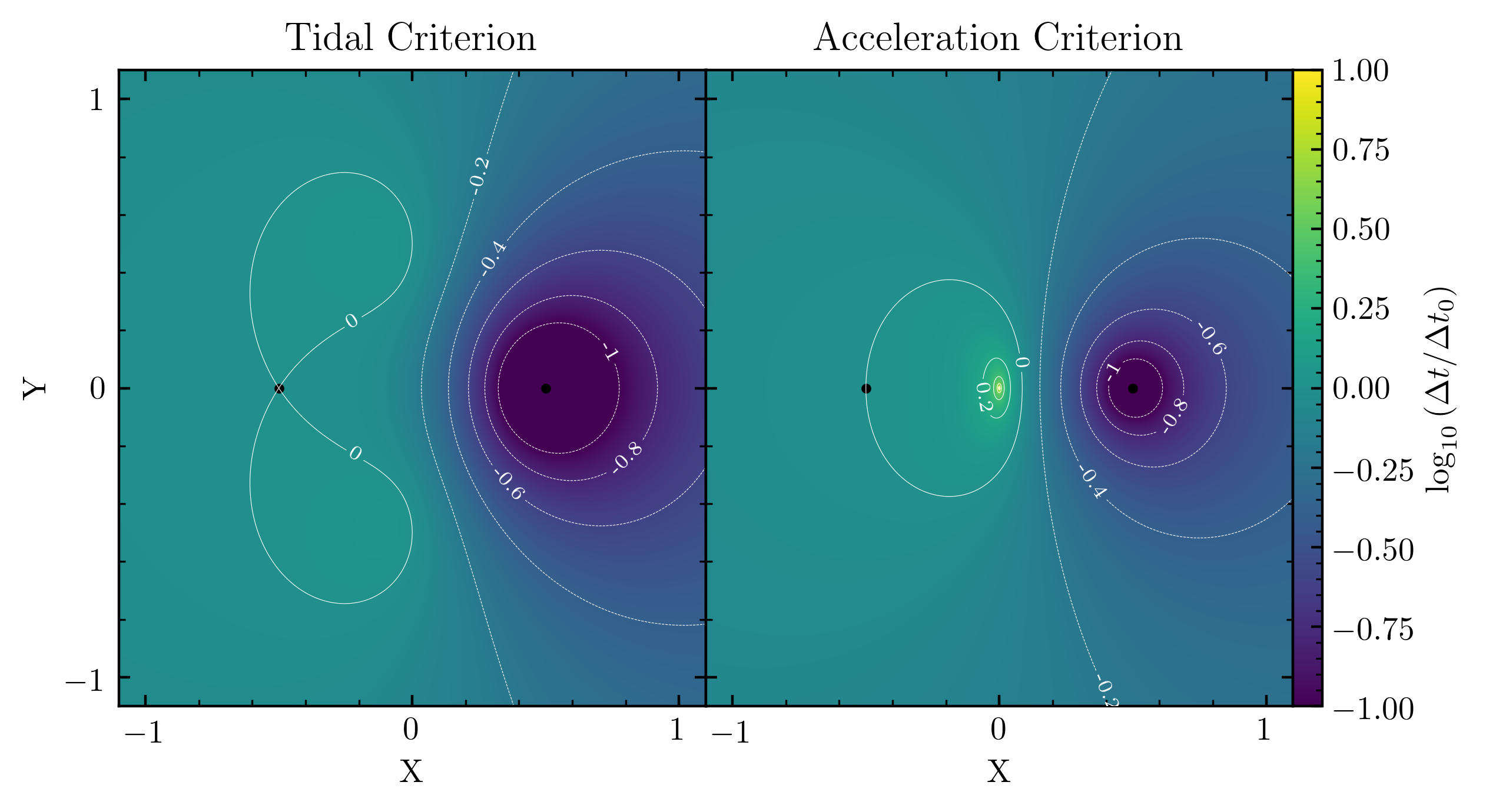}
    \caption{Effect of adding a perturber upon the timestep taken, for our proposed tidal timestep criterion (Eq.~\ref{eq:dt_tidal}; left) and the commonly-adopted acceleration criterion (Eq.~\ref{eq:power}; right). 
    We plot the timestep $\Delta t$ obtained (colors) at locations $(x,y)$ in the potential of a pair of equal point-masses (black points at $(\pm0.5,0)$), relative to the timestep $\Delta t_0$ obtained if only the point-mass at $x=-0.5$ were present. 
    Adding a second mass makes the tidal timestep (Eq.~\ref{eq:dt_tidal}) stricter than before essentially everywhere ($\Delta t \lesssim \Delta t_0$). Near the Lagrange point at $(0,0)$, the acceleration vanishes and so Eq.~\ref{eq:power} allows $\Delta t \gg \Delta t_0$, and confines the region of small timesteps near $(+0.5,0.5)$ to a smaller volume.}
    \label{fig:contours}
\end{figure*}
We now explore how the tidal timestep criterion behaves in a wider range of gravitational potentials. First, consider how the tidal and acceleration criterion behave in the simplest non-trivial case: in the vicinity of two point masses. In Figure \ref{fig:contours} we plot the local timestep in this potential, relative to the value it would have if only one of the point masses were present. The acceleration criterion breaks down at the midpoint between the two masses, reaching arbitrarily large values as $\mathbf{a}\rightarrow 0$. With the tidal criterion, the second mass makes the timestep stricter than before everywhere except within a small region where it is slightly ($< 1.07$ times) larger. Thus the tidal criterion has the desirable property that adding additional perturbations to the system typically results in a stricter timestep. This is essentially because only very special mass distributions can cause the norm of the tidal tensor to cancel out entirely (e.g. inside a spherical shell, where a large gravitational timestep is physically correct).

Now consider a spherically-symmetric mass distribution whose central region is described by a power-law:
\begin{equation}
    \rho\left(r\right) = \rho_0 \left(\frac{r}{r_0}\right)^\alpha.
    \label{eq:powerlaw}
\end{equation}
The tidal tensor at a point on the x-axis is:
\begin{equation}
    \mathbf{T} = -\frac{G M\left(<r\right)}{r^3} 
\begin{bmatrix}
    \alpha+1 & 0 & 0 \\
    0 & 1 & 0 \\ 
    0 & 0 & 1 
\end{bmatrix}
\end{equation}
and we have
\begin{equation}
    t_{\rm tidal}=\left(\frac{6}{3+2\alpha+\alpha^2}\right)^{1/4} t_{\rm dyn}, 
\end{equation}
where for $-3 < \alpha < -1$ we have $t_{\rm tidal}$ between $1-1.3\,t_{\rm dyn}$ (with $t_{\rm dyn} \equiv \Omega^{-1} = (G\,M(<r)/r^{3})^{-1/2}$). Thus $t_\mathrm{tidal}$ recovers the desired Keplerian form as $r\rightarrow \infty$ and always closely reflects the local dynamical time.
It is worth noting that the correct $t_{\rm dyn}$ is recovered by $t_{\rm tidal}$ even in models which do not have finite mass: e.g.\ the singular isothermal sphere ($\alpha=-2$ above), or a \citet{nfw:profile} profile (with $\alpha\rightarrow -1$ and $t_{\rm tidal} \approx 1.3\,\Omega^{-1}$ as $r\rightarrow 0$, and $\alpha \rightarrow -3$ and $t_{\rm tidal} \approx \Omega^{-1}$ as $r\rightarrow \infty$).


\subsection{Implementation}

The most efficient method for computing $\mathbf{T}$ in simulations depends on the algorithm used to compute ${\bf g}$. In many N-body methods, the most straightforward method is to use the linearity of $\mathbf{T}$ to sum the contributions of each particle:
\begin{equation}
    T_{ij}\left(\mathbf{x}\right) = \sum_{n=0}^{N} -\partial_i \partial_j \Phi_n\left(\mathbf{x} - \mathbf{x}_n\right) = \sum_{n=0}^{N} \partial_i \left[ {\bf g}_{n,\,j} \left(\mathbf{x} - \mathbf{x}_n\right) \right] ,
    \label{eq:tij_sum}
\end{equation}
where $\Phi_n$ (or ${\bf g}_{n\,j}$) is the contribution of particle $n$ to the gravitational potential (or acceleration component $j$) as a function of the separation $\mathbf{x} - \mathbf{x}_n$, and the partial derivatives are with respect to the target position $\mathbf{x}$. For point masses this amounts to summing Eq.~\ref{eq:tij_kep}, however if a softened force law is used for gravity then the tidal contribution should be softened accordingly. The function $-\partial_i \partial_j \Phi_n \left( \mathbf{x} - \mathbf{x}_n \right) = \partial_i \left[ {\bf g}_{n\,j} \left(\mathbf{x} - \mathbf{x}_n\right) \right]$ is typically known analytically for all ${\bf x}-{\bf x}_{n}$, so this can in principle be evaluated in the same particle sweep as the evaluation of ${\bf g}$ in e.g.\ tree or direct N-body methods. 

In Fourier-based methods, where Fourier transforms are used to solve the Poisson equation for $\Phi$ and then transformed back to obtain ${\bf g}$ (after replacing $\partial_j$ with wavenumber $k_j$ in Fourier space), evaluating $\mathbf{T}$ is also straightforward, as it simply ``pulls down'' an additional power of $k$. 

Lastly, if the gravitational potential or the gravitational field is known at the position of each resolution element in the simulation, $\mathbf{T}$ can be estimated via direct numerical differentiation of ${\bf g}({\bf x}_{i})$ or $\Phi({\bf x}_{i})$, in a gradient sweep.

\section{Numerical Tests}
\label{section:tests}
\begin{figure}
    \centering
    \includegraphics[width=0.45\textwidth]{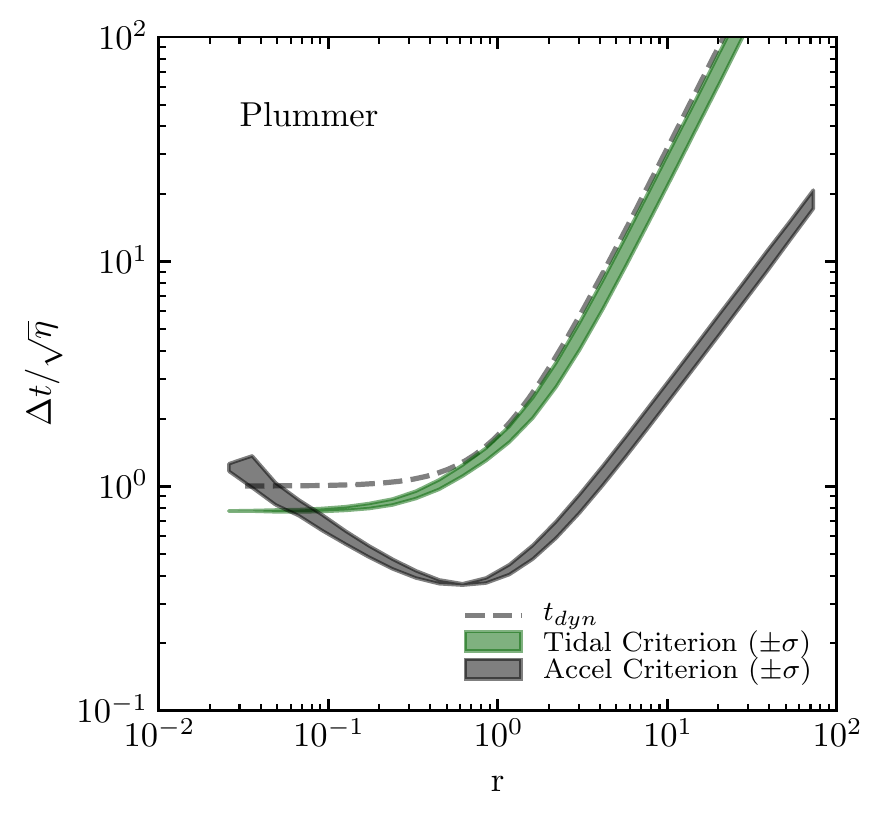} \includegraphics[width=0.45\textwidth]{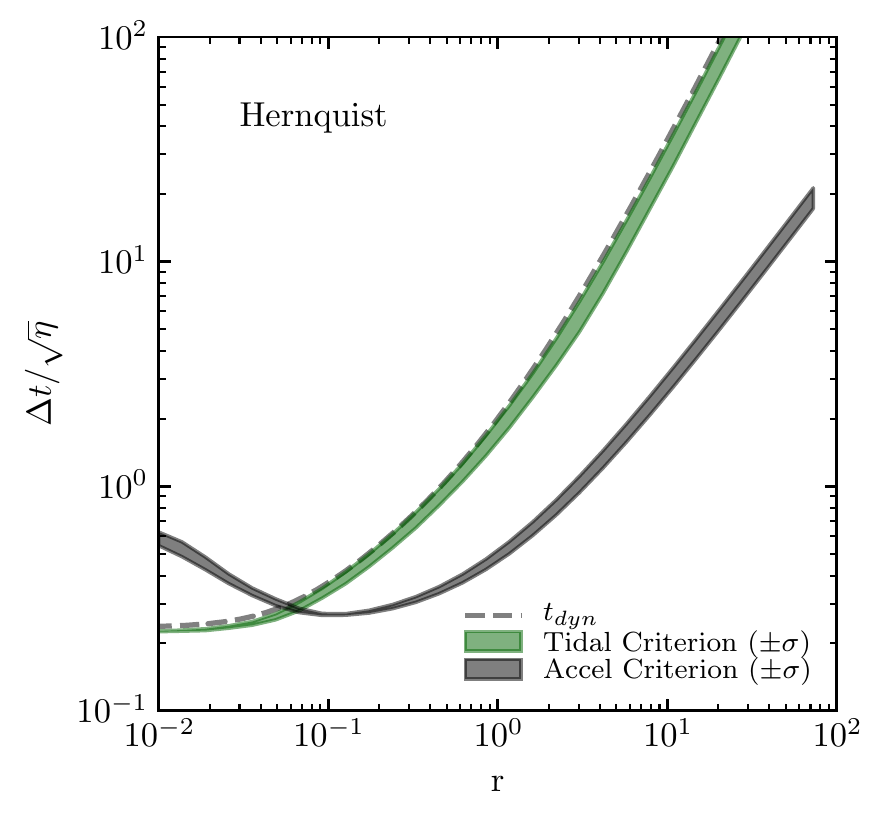}
    \caption{Timesteps calculated for each particle in a snapshot of a Plummer model (top; with softening $\epsilon=0.1$) and Hernquist model (bottom; $\epsilon=0.05$) in collisionless equilibrium with $G=M=a=1$ and $10^{5}$ particles, scaled to the tolerance parameter $\sqrt{\eta}$. Contours indicate $\pm \sigma$ quantiles of particles' timesteps in radial bins. We compare the tidal (Eq.~\ref{eq:dt_tidal}) and acceleration (Eq.~\ref{eq:power}) criteria, as well as the ``dynamical time'' $t_{\rm dyn} \equiv (G M(<r)/r^{3})^{-1/2}$. The tidal criterion generally tracks $t_{\rm dyn}$ more closely and does not diverge as $r\rightarrow0$.}
    \label{fig:dt_vs_R}
\end{figure}

\begin{figure}
    \centering
    \includegraphics[width=\columnwidth]{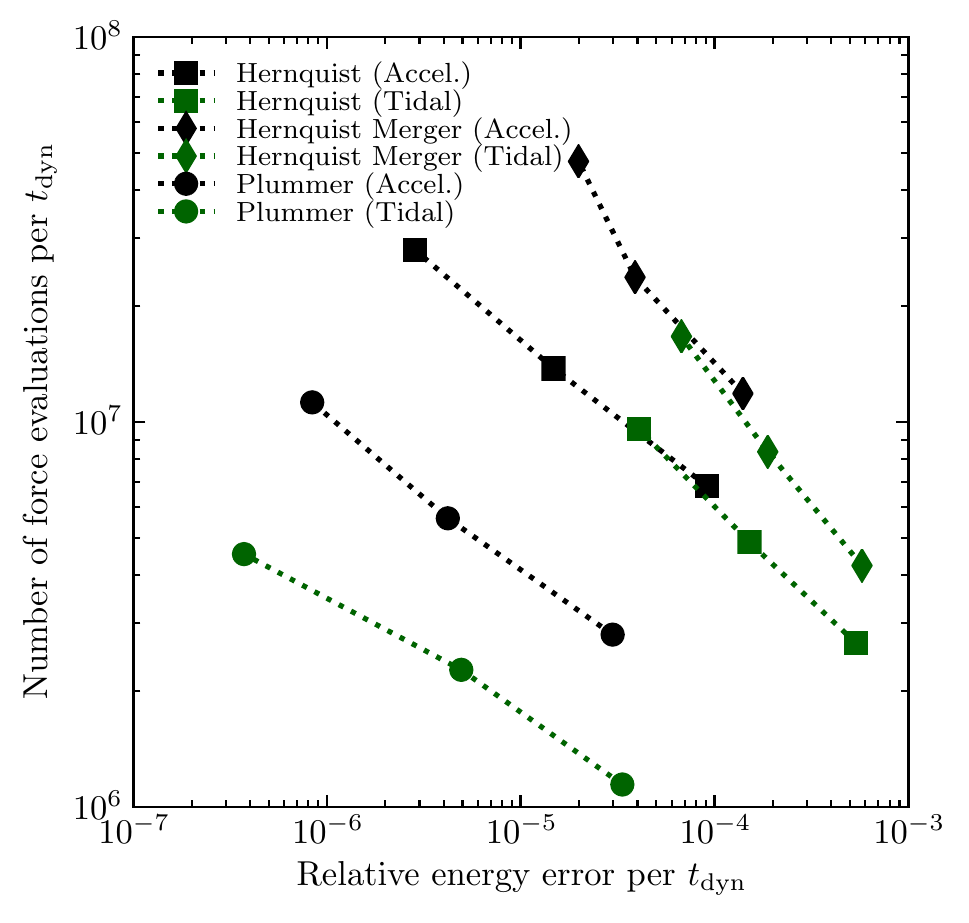}
    \caption{Relative energy error per $t_\mathrm{dyn}$ as a function of number of force evaluations per $t_\mathrm{dyn}$ in different simulations, using the tidal criterion (Eq. \ref{eq:dt_tidal}) and the acceleration criterion (Eq. \ref{eq:power}) with $\eta=0.01$, $0.0025$, and $0.00625$ as implemented in {\small GIZMO}.  This isolates the respective efficiencies of the timestepping schemes, disregarding additional overheads incurred by computing the tidal tensor. In the isolated and merging Hernquist problems, the tidal criterion is as efficient as the acceleration criterion, while in the isolated Plummer sphere it is much more efficient, owing to the fact that the central acceleration vanishes (so Eq. \ref{eq:power} breaks down).}
    \label{fig:Nfev}
\end{figure}

\begin{figure}
    \centering
    \includegraphics[width=\columnwidth]{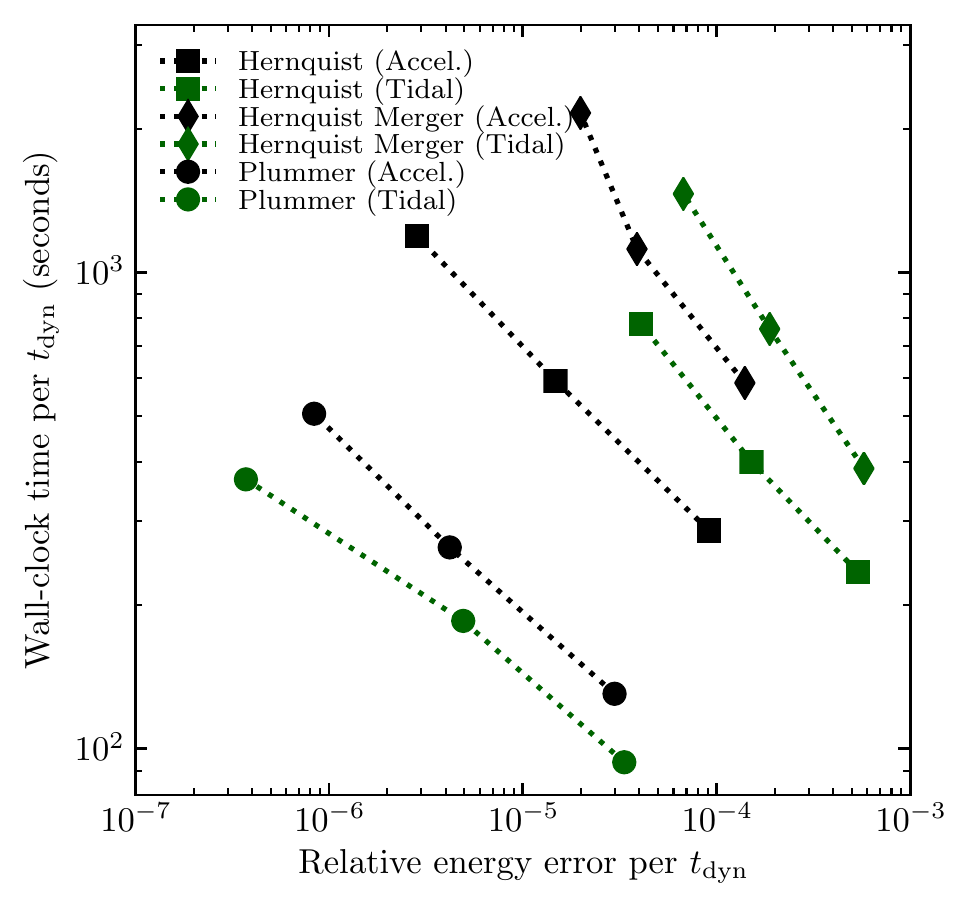}
    \caption{Relative error per $t_\mathrm{dyn}$ as a function of actual wall-clock time per $t_\mathrm{dyn}$ in different simulations, using the tidal and acceleration criteria with $\eta=0.01$, $0.0025$, and $0.00625$. The relative overhead of the tidal tensor computation (a factor of $\sim 2$ in the tree-force computation) makes the tidal criterion somewhat less efficient in the Hernquist problems, however it is superior in the Plummer problem despite the overhead.}
    \label{fig:wallclock}
\end{figure}

We will now test the efficiency of the tidal timestep criterion given by Eq.~\ref{eq:dt_tidal} in N-body simulations. For this purpose, different measures of efficiency can be defined. As is standard, we will use the relative energy error $|\Delta E|/E$ to quantify the level of integration error.
For a code using an integration scheme of $n$'th order, the energy error is expected to scale $\propto N_{\rm step}^{-n}$, where $N_{\rm step}$ is the number of particle-timesteps taken in the simulation. Therefore, neglecting other sources of error such as force errors, the total number of particle-timesteps to achieve a given energy error will generally scale as
\begin{equation}
    N_{\rm step} \propto \left(\frac{|\Delta E|}{E}\right)^{-1/n}.
\end{equation}
As the integration accuracy can always be adjusted by varying the timestepping tolerance parameter $\eta$, one figure of merit for a timestep criterion is the {\it total number of timesteps required for a solution of a given energy error}.

This definition of efficiency quantifies the raw efficiency of the method as a timestepping scheme, neglecting any computational overheads. However, in the case of the tidal criterion, we shall find that the additional computational overhead of the tidal tensor calculation increases the CPU time per particle-timestep. Therefore, a complete evaluation should consider the total {\it wall-clock time to solution} required for a given accuracy. We will evaluate both notions of efficiency.

Because we simulate only gravitational physics, our results represent the worst-case scenario for the overall computational efficiency of the tidal criterion in comparison to the acceleration criterion. In general, the relative overhead of the tidal tensor summation will be reduced by the addition of any other physics modules beyond just gravity.

\subsection{Code}
We run simulations with {\small GIZMO} \citep{hopkins:gizmo}, a multi-method, multi-physics N-body and hydrodynamics simulation code. The subset of {\small GIZMO}'s N-body integration, domain decomposition, and gravity solver routines are largely derived from {\small GADGET-3}, itself descended from {\small GADGET-2} \citep{springel:gadget}. Gravity is solved via the approximate \citet{barneshut} tree-code method using a geometric opening criterion $L_{\rm node}/R < 0.5$, where $L_{\rm node}$ is the side-length of an octree node and $R$ is the distance between the target position and the node centre of mass. The equations of motion are integrated according to the second-order `kick-drift-kick' (KDK) scheme with an adaptive timestep \citep{springel:gadget}. Timesteps are discretized in a powers-of-2 hierarchy, where the timestep taken by a particle is the largest power-of-two subdivision of the total simulation time that satisfies the timestep criteria. For a given simulation, we adopt a fixed gravitational softening according to a standard cubic-spline kernel (see \citealt{hopkins:gizmo}). Throughout, we set the softenings roughly according to the `optimal' values given in \citet{dehnen:2001.softening}, minimizing the mean force error with respect to the underlying collisionless equilibrium model. When quoting gravitational softening lengths in this work, we quote the Plummer-equivalent softening length $\epsilon$, a factor of $2.8$ smaller than that radius of support of the cubic spline kernel. We sum the tidal tensor at each position in the simulation in the same pass through the Barnes-Hut tree as is used to sum the gravitational field, and use the same gravitational softening. This is essentially the same as the method used to calculate $\mathbf{T}$ for cosmological applications in  \citet{vogelsberger:2008.gde}.

\subsection{Simulation Setup}
We run three different collisonless N-body problems, to evaluate how the tidal and acceleration criteria perform when integrating paticle motion in different types of potentials. In all simulations $G=1$.

First, we simulate an isotropic \citet{plummer} model with unit mass and scale radius, described by the density distribution
\begin{equation}
    \rho \left(r\right) = \frac{3}{4\pi}\frac{1}{(1+r^2)^{5/2}}.
\end{equation}
We initialize the model in collisionless equilibrium with $10^5$ particles, and run it for 1000 time units. The Plummer-equivalent gravitational softening length is set to $0.1$. This model has a flat inner density profile and a relatively steep fall-off in density, and thus roughly models gravitationally-bound star clusters.

Second, we simulate a \citet{hernquist:model} with unit mass and scale radius, described by the density distribution
\begin{equation}
    \rho\left(r\right) = \frac{1}{2\pi}\frac{1}{r (1+r)^3}.
\end{equation}
We also initialize this model in collisionless equilibrium with $10^5$ particles and simulate it for $1000$ time units. The Plummer-equivalent gravitational force softening is set to 0.05. This model has an inner density cusp of the form $\rho \propto r^{-1}$, and thus serves as a model for the collisionless (ie. bulge or dark matter halo) component of a galaxy, where such density cusps form naturally \citep{nfw:profile}.

Lastly, we simulate a merger between two copies of the above Hernquist model. We initialized the models 10 distance units apart on a parabolic encounter with a pericenter of 1. The encounter was then simulated for 100 time units, during which the galaxies eventually merge after making an initial close passage. This problem probes how the timestep criteria perform in a more dynamic situation, when the gravitational potential is varying on a timescale comparable to the dynamical time and the system undergoes violent relaxation.

\subsection{Results}
\subsubsection{Numerically-computed timesteps}
In Figure \ref{fig:dt_vs_R} we plot the timesteps calculated by the code for each particle in the isolated Plummer and Hernquist simulations as a function of radius, in the initial snapshot of each respectively simulation. As expected, the timestep set according to the tidal criterion tracks the true dynamical time more closely in models where it is well-defined. The acceleration criterion generally computes a timestep that reaches a minimum at a radius on the order of the turnover in the system's density profile, rising as $r \rightarrow 0$ and as $r\rightarrow \infty$, but more slowly (a factor of $r^{-1/2}$) than the dynamical time. In the Plummer model, we also see that the acceleration criterion computes a timestep that begins to rise toward the centre of the cluster, because the gravitational field approaches zero. In contrast, the tidal timestep generally decreases monotonically as one approaches the centre of the system.

\subsubsection{Energy error sign difference}
\label{section:energysign}
The total energy in the simulations evolved due to numerical errors, as expected. We have verified that these errors are dominated by integration error, as opposed to errors from the approximate Barnes-Hut force scheme. For both timestep criteria, the error exhibits a secular trend $|\Delta E| \propto t$, as opposed to oscillating about some mean value, since our KDK integration scheme is not exactly symplectic when individual timesteps are used \citep{springel:gadget}. For the acceleration criterion, the energy always {\it increased} over time (ie. integration errors induced numerical heating), whereas for the tidal criterion, the energy {\it decreased} (ie. numerical cooling). Thus in the limit where errors are large, we expect systems evolved with the acceleration criterion to artificially puff up over time due to integration errors. Conversely, they will grow artificially denser when evolved according to tidal criterion.

\subsubsection{Energy error magnitude and efficiency}
The efficiencies of the different timestep prescriptions are demonstrated in Figure \ref{fig:Nfev}, in which we plot the relative energy error per $t_\mathrm{dyn}$ as a function of the number of force evaluations per $t_\mathrm{dyn}$, using the half-mass $t_\mathrm{dyn}$ for each respective model. In all problems, the tidal criterion requires at most roughly the same number of force evaluations for a given energy accuracy. In the Hernquist problems, the performance is roughly comparable.\footnote{It should be noted that in the Hernquist problems, the two timestep criteria have about the same efficiency, but as we have defined them they do not achieve the same error for the identical value of $\eta$. This is merely a question of our (arbitrary) normalization convention. If we take the Hernquist model as the ``worst-case'' scenario, then achieving comparable errors and/or the same average number of timesteps/force evaluations requires setting $\eta$ in Eq.~\ref{eq:dt_tidal} (the tidal criterion) a factor $\sim 1/2$ lower than the $\eta$ used in Eq.~\ref{eq:power} (the acceleration criterion).} However the tidal criterion is much more efficient in the Plummer problem, incurring an order of magnitude less energy error for a given number of force evaluations. We believe that this is due to the stricter timestep in the centre of the cluster, where the dynamical timescale is the shortest and errors can be accumulated over many shorter orbits. The tidal criterion is stricter here, while the acceleration criterion breaks down (see Figure \ref{fig:dt_vs_R}).

Figure \ref{fig:wallclock} compares the errors versus  wall-clock time, where the extra overhead of the tidal tensor calculation makes the tidal criterion somewhat less efficient than the acceleration criterion in the Hernquist problem. But the tidal criterion is still more efficient in the Plummer problem (despite the overhead) owing to its much higher efficiency per-force-evaluation. Of course, the cost in wall-clock time will depend on the particular code, implementation of gravity and details of the tidal tensor calculation, and whether other physics is included.

\subsubsection{Density profiles}
\label{sec:massprofile} 

\begin{figure}
    \includegraphics[width=\columnwidth]{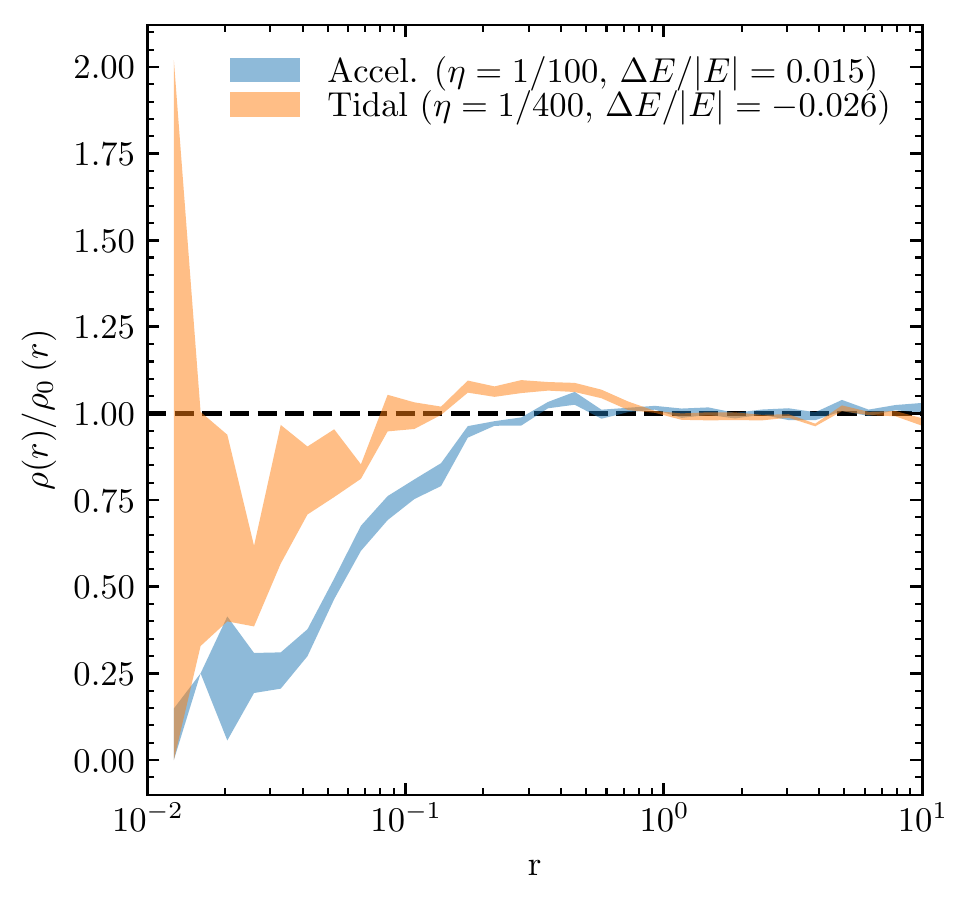}
    \caption{Radial mass density profiles after 1000 time units ($\sim 200 t_\mathrm{dyn}$) for the Hernquist model, evolved according to the acceleration (blue) and tidal (orange) timestep criteria, scaled relative to the density profile of the initial collisionless equilibrium $\rho_\mathrm{0}\left(r\right)$. Shaded intervals represent the $\pm \sigma$ quantiles over the final 5 snapshots.} 
    \label{fig:massprofile}
\end{figure}

The energy error is a useful heuristic for the accuracy of an N-body simulation, but it is also important to analyze other metrics, such as the evolution of the density profile according to the different integration schemes. For this analysis we focus on the Hernquist model, because its dense central cusp region evolves more rapidly than the core of the Plummer model, making any purely-numerical effects more pronounced.

In Figure \ref{fig:massprofile} we compare the density profiles in our models after 1000 time units ($\sim 200$ half-mass dynamical times) for our isolated Hernquist models, relative to the initial density profile. We use runs with comparable energy error magnitude for comparison: the $\eta=0.01$ and $\eta=0.0025$ for the acceleration and tidal criteria respectively. Our initial conditions generator did not account for softening \citep[e.g.][]{barnes:2012.softening.is.smoothing}, so the distribution quickly relaxes from the exact, un-softened Hernquist model toward a new collisionless equilibrium that deviates from the central $\propto r^{-1}$ profile below radii comparable to the softening scale. We use this new equilibrium to define our ``initial" profile, by taking the median density in radial bins for 5 snapshots between times 10 and 20 (ie. 2-4$t_\mathrm{dyn}$). We plot the $\pm \sigma$ intervals of the density profiles of the final 5 snapshots, scaled relative to this initial profile. 

After 1000 time units, the density profiles have evolved to be statistically distinguishable from the initial profile within $r \sim 1$. We find that the density profile obtained with the tidal criterion is generally greater than that obtained with the acceleration criterion. This is due to the schemes' opposite respective energy errors: the numerical cooling inherent in the tidal scheme (\S \ref{section:energysign}) should cause the system to grow denser in accord with the virial theorem, whereas numerical heating inherent in the acceleration criterion may cause the central region to be evacuated as the central particles move to higher orbits. The acceleration criterion systematically under-predicts the density within $r\sim 0.5$, eventually by an order of magnitude at $r=0.01$. The tidal criterion {\it overpredicts} the density from $r=0.1-0.5$, and underpredicts it by a factor of order unity at all resolvable radii within $r<0.1$, likely due to numerical relaxation.


\section{Discussion}
\subsection{Cosmological simulations}
We have also benchmarked the tidal criterion against the acceleration criterion in dark matter-only cosmological simulations, essentially repeating the experiments performed in \citet{power:2003}. In these simulations, where dark matter halos with \citet{nfw:profile}-like density profiles form during gravitational collapse, artificial heating due to integration errors in the centres of dark matter halos can lead to an artificially-large core radius, and thus numerical convergence will be hampered. This error is in fact identical in nature to the energy errors accumulated in the core of the \citet{hernquist:model} model (which is identical to NFW in the core) in the previous section. Unsurprisingly, we find very similar performance and convergence behaviour for the tidal criterion and acceleration criterion at the same number of force evaluations (as we did for the \citet{hernquist:model} models). 

\subsection{``Multi-Physics'' Simulations}
We have also begun to explore application of Eq.~\ref{eq:dt_tidal} in multi-physics simulations of star cluster formation (following e.g.\ \citealt{grudic:2016.sfe}) and galaxy formation (following \citealt{fire2}). These simulations include self-gravity of gas and collisionless particles (stars, dark matter), radiation-magneto-hydrodynamics, chemistry and radiative cooling, dynamical on-the-fly formation of new sinks or star particles, cosmological integration, and other ``sub-grid'' physics (e.g.\ feedback from newly-formed stars). The actual simulations will be presented and studied in future work, but we briefly describe the effects of the timestep criterion here. Note that in these cases, the multi-physics nature means that there are many different timestep criteria checked (with the shortest being the one used by the code), so it is difficult to always identify an unambiguous effect of Eq.~\ref{eq:dt_tidal}. We generally find that integration errors in the gravitational forces become sub-dominant to other sources of error (e.g.\ hydrodynamics) for modest $\eta \ll 0.1$. In all these cases, the computational overhead of computing $\mathbf{T}$ is negligible.

In our previous studies of star cluster formation using the acceleration criterion (e.g.\ \citealt{guszejnov:2018.isothermal}), we found that if we used adaptive force-softening (setting $\epsilon$ to the inter-particle separation), this artificially softened and suppressed what should be ``true'' binaries and multiples. But using a small (or zero) force softening to represent point masses (following e.g.\ \citealt{bate:2009.imf}) with Eq.~\ref{eq:power} leads to arbitrarily small timesteps in the limit where $\epsilon$ ceases to correspond to any physical time-scale in the problem. Eq.~\ref{eq:dt_tidal} more accurately and efficiently treats these cases, and we have confirmed that it simultaneously accurately recovers the bulk orbital timescale of stars in the cluster when close encounters are not important (the collisionless limit) but also correctly switches to being dominated by the tidal contribution of the nearest star (and accurately integrating multiples) when close encounters occur. In our galaxy formation simulations, other uncertain physics (e.g.\ stellar ``feedback'' processes) strongly dominate the error budget so the effects of improved integration errors are generally minimal. However, when dense, gravitationally-bound star clusters formed in these runs \citep[studied in][]{kim:2018.fire.gcs, ma:2019.fire.gcs}, they generally had Plummer-like density profiles which would relatively quickly ``dissolve'' owing to the accumulation of integration errors when the acceleration criterion was used (importantly, in a galaxy-scale simulation, such clusters are not very well-resolved). Using Eq.~\ref{eq:dt_tidal}, the superior energy conservation in Plummer-like potentials slows down this process substantially, giving order-of-magnitude longer dissolution times for marginally-resolved clusters.

\section{Conclusions} 
We propose and test a new timestep criterion for integration of gravitational motion (Eq.~\ref{eq:dt_tidal}), scaled to the characteristic timescale defined by the tidal tensor (the timescale over which the gravitational accelerations can change appreciably). We show that the tidal criterion performs at least as well as the commonly-adopted acceleration-based criterion (Eq. \ref{eq:power}), in terms of number of particle-timesteps required to achieve a desired integration accuracy. When integrating orbits in potentials where the acceleration vanishes or becomes small anywhere (e.g. the center of a Plummer or constant-density sphere), the tidal scheme has notably superior performance, more accurately integrating orbits through central passages where the acceleration criterion breaks down. Moreover the tidal criterion is independent of any uniform boost or acceleration (respects the equivalence principle), is coordinate-independent, depends only on quantities needed for the computation of gravitational forces (so can generally be evaluated in the same sweep used to compute accelerations), is independent of the force softening on scales larger than that softening, does not make reference to any assumptions about the matter distribution or potential structure, and has a straightforward physical motivation.

There is a vast body of work describing specialized integration methods for evolving collisional N-body systems \citep[see][and references therein]{aarseth:2003}, and we emphasize that specialized methods designed for a particular problem will almost always outperform a general-purpose criterion like Eq.~\ref{eq:dt_tidal}. The role of a general-purpose timestep criterion is not to be optimal in all cases but rather to be robust, ensuring that an acceptable solution will be obtained in a wide range of problems with as little fine-tuning as possible. In this respect, Eq.~\ref{eq:dt_tidal} performs well, and most importantly should interpolate correctly ``on the fly'' between ``collisionless'' limits (where a given particle is moving through a smooth background potential) and ``collisional'' limits (e.g.\ true hard binaries, resolved clusters), without any fine-tuning by the user. 
For broad classes of simulations of e.g.\ planet star and galaxy formation, N-body dynamics, and self-gravitating fluid dynamics, this should be useful.

\section*{Acknowledgements}
We thank D\'{a}vid Guszejnov, Walter Dehnen, Oliver Hahn, and Jens St\"{u}cker for enlightening discussions, and thank the anonymous referee for helpful suggestions that improved the text. Support for MYG was provided by a CIERA Postdoctoral Fellowship. Support for MYG and PFH was provided by an Alfred P. Sloan Research Fellowship, NSF Collaborative Research Grant \#1715847 and CAREER grant \#1455342, and NASA grants NNX15AT06G, JPL 1589742, 17-ATP17-0214. Numerical calculations were run on the Caltech compute cluster ``Wheeler,''. This research has made use of use of NASA's Astrophysics Data System, {\texttt ipython} \citep{ipython}, {\texttt numpy}, {\texttt scipy} \citep{scipy}, and {\texttt matplotlib} \citep{matplotlib}.



\bibliographystyle{mnras}
\bibliography{master} 






\bsp	
\label{lastpage}
\end{document}